\documentclass[twocolumn]{IEEEtran}

\newcommand{\BibTeX}{{\rm B\kern-.05em{\sc i\kern-.025em b}\kern-.08em
    T\kern-.1667em\lower.7ex\hbox{E}\kern-.125emX}}

\usepackage{graphicx}
\usepackage{amsmath}

\title{Power Consumption of LDPC Decoders in Software Radio}

\author{Chia-han Lee and Wayne Wolf}

\begin{document}

\maketitle

\begin{abstract} 

LDPC code is a powerful error correcting code and has been applied to many advanced communication systems.
The prosperity of software radio has motivated us to investigate the implementation of LDPC decoders on processors.
In this paper, we estimate and compare complexity and power consumption of LDPC decoding algorithms running on general purpose processors.
Using the estimation results, we show two power control schemes for software radio: SNR-based algorithm diversity and joint transmit power and receiver energy management.
Overall, this paper discusses general concerns about using processors as the software radio platform for the implementation of LDPC decoders.

\end{abstract}

\section{Introduction}

Software radio is an advanced radio communication system that can process any modulation scheme and coding scheme at various frequency bands by means of as little hardware as possible, processing the digitized signals in software. 
Since the introduction by Mitola in the early 90's~\cite{Mitola:1993}, software radio has become an active research paradigm. 
Software radio originated from the global demand to develop a transceiver which can easily switch between multiple standards and recognize different signal types. 
For military purpose, a software radio system can process hetero\-geneous signals transmitted from different generations of equipments used by different troops~\cite{Lackey:1995}. 
Moreover, flexible radio avoids interception or jamming from the enemy. 
As a consumer product, a software radio transceiver serves as a universal cell phone when traveling around the world. 
Some advanced communication architectures allow a radio to be changed on-the-fly (frequency band, modulation scheme, error correction, etc.) according to the radio environments~\cite{Lee:2006-arch}. 
Utilizing different kinds of modulation and multiple frequency bands allows more efficient use of precious radio spectrum~\cite{Haykin:2005}. 
As a result, software radio is getting more and more important due to the world trend toward multi-standard and multi-service communications.
Traditional communication systems use different hardware for different components of a communication system.
For example, computationally demanding tasks are implemented on application-specific integrated circuits (ASIC) while less demanding ones are run on digital signal processors (DSP). 
Accelerators are implemented on ASIC or field-programmable gate array (FPGA), and applications or controllers are usually run on DSPs or microprocessors.
In contrast to the traditional realization of filtering and up- and down-conversion in analog domain, a communication system realized in software requires almost all the functions to be implemented on general purpose processors (GPP) or digital signal processors.
This is done by placing analog-to-digital converter (ADC) as close as possible to the antenna, ideally right after the low noise amplifier (LNA) and bandpass filter (BPF).
The remaining radio frequency (RF) and intermediate frequency (IF) functions, such as channelization, downconversion, synchronization, and filtering, are performed in digital domain on general purpose processors~\cite{Lee:2006-arch}. 
Unfortunately, performing high-frequency and high-data rate functions in software requires significant amount of computation. 
Power consumption and delay issues must therefore be investigated and solved in a software radio system.

Error correcting coding is a critical part in today's communication system.
Nevertheless, the complexity of error correcting codes is considerably high compared to other parts of the communication system.
Error correcting codes, therefore, must be carefully inspected and designed in software radio scenario.
Advanced error correcting codes, such as low-density parity-check (LDPC) codes, yield very good error correcting capability but they are also hardware demanding and power hungry.
Understanding how LDPC codes perform on processor-based software radio platform helps us find new directions and new solutions for LDPC decoding.

Power is an important concern, especially in a software radio system which uses general purpose processors as its hardware platform. 
Unfortunately, no previous work has ever focused on the comparison of energy/power consumption of different LDPC decoding algorithms on general purpose processors, although some did present the power consumption analysis for their own decoder architecture~\cite{Mansour:2002}. 
That motivates our work of analyzing decoding algorithms and proposing an efficient methodology to estimate the power consumption.
In this paper, we will discuss several popular LDPC decoding algorithms and their realization in software radio.
However, it is not our purpose to have a thorough comparison of all decoding algorithms.
Some of the newer decoding schemes might not be included, but the analysis process can be easily extended to apply to those algorithms too.

The organization of this paper is the following. Section~\ref{ldpc_code} gives an introduction to the LDPC codes. Decoding algorithms are described, and their complexity and performance are compared. Section~\ref{processor_platform} discusses concerns about implementation of LDPC codes on processors, including algorithm simplification, power consumption estimation, throughput estimation, and cache behavior analysis. In Section~\ref{power_control}, we introduce two efficient power control schemes based on algorithm diversity for software radio.

\section{LDPC Codes}\label{ldpc_code}

\subsection{Brief Introduction}

\begin{figure}[!t]
  \begin{center}
  \includegraphics[width=3.5in]{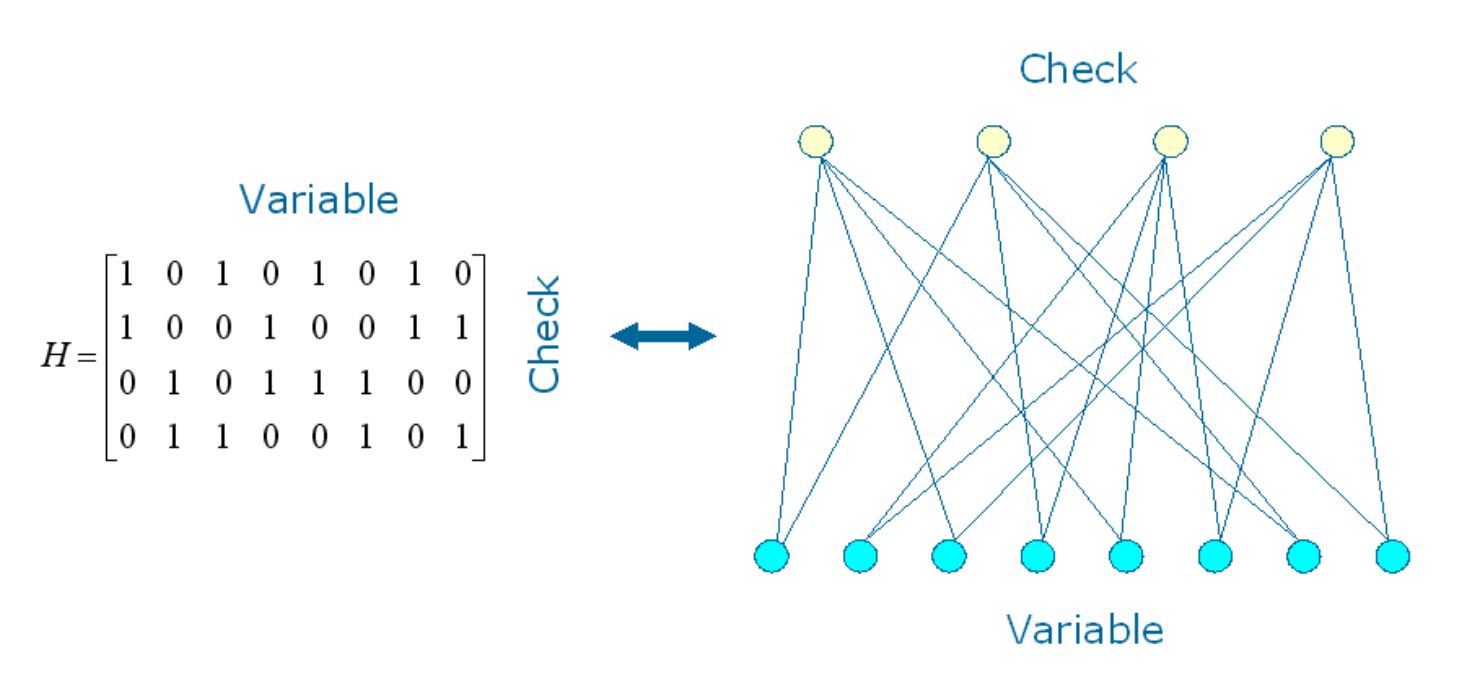}
  \end{center}
  \caption{Mapping between a parity-check matrix and Tanner graph of a (8, 4) LDPC code.}
  \label{fig-tanner}
\end{figure}

An LDPC code is defined by its parity-check matrix $H$. 
As revealed from its name, the parity-check matrix of an LDPC code is "low density", which means only very few entries in the matrix are nonzero. 
Not only does the parity-check matrix define a LDPC code, but it also provides a convenient view of iterative decoding process when mapped to Tanner graph. 
Tanner graph~\cite{Tanner:1981} is a bipartite graph with variable nodes (or bit nodes) on one side and check nodes on the other side.
Each of the variable nodes corresponds to a bit in the codeword, and each of the check nodes sets one parity-check constraint through the edges connecting variable and check nodes in a Tanner graph. 
Variable nodes, check nodes, and edges in a Tanner graph can be directly mapped to/from a parity-check matrix. 
Every column in the parity-check matrix corresponds to a variable node and every row maps to a check node.
Each entry in the parity-check matrix reveals whether there is an edge between particular variable and check node pair. 
Fig.~\ref{fig-tanner} shows an example of mapping from a parity-check matrix to a Tanner graph. 
The Tanner graph clearly presents how parity-check bits set constraints on information bits, so it is widely used for describing the concept of iterative decoding or message-passing algorithms.

\subsection{Decoding of LDPC Codes}

The decoding of LDPC codes uses iterative decoding and is easy to understand with the help of Tanner graph. 
The idea of iterative decoding is to pass messages between variable and check nodes to update the information of the reliability of received bits. 
What kind of messages needed to be passed depends on which decoding algorithm is used. 
At variables nodes, messages from channel bits and check nodes are combined and then passed to check nodes. 
Similarly at check nodes, messages from variable nodes are combined and then passed back to variable nodes.
When messages are passed from variable nodes to check nodes and then back to variable nodes again, it is called one iteration. 
The iterative process is repeated until all bits corrupted by channel are corrected (the decoded vector is a codeword, i.e. $H_d^T = 0$, where $d$ is the decoded vector) or a predetermined number of iteration is reached.
Iterative decoding of LDPC codes falls into two main categories: belief propagation (BP)-based and bit-flipping(BF)-based. 
The most well-known BP-based algorithms are sum-product (SP)~\cite{Kschischang:2001}~\cite{Pearl:1988} and its logarithm domain variant: log sum-product (log-SP)~\cite{Chiani:2000}~\cite{ZhangT:2004}. 
A significant portion of the computation load in these two algorithms is on the $tanh$ function.
A much simpler algorithm called min-sum (MS)~\cite{Wiber:1996} and its modified version, modified min-sum (MMS)~\cite{Heo:2003}, remove the need of this function such that the computation burden can be alleviated. 
Bit-flipping-based algorithms, on the other hand, are based on flipping the least unreliable bit after each iteration. 
Weighted bit-flipping (WBF)~\cite{Kou:2001} is a modified version of the original bit-flipping algorithm proposed by Gallager~\cite{Gallager:1962}. 
Modified weighted bit-flipping (MWBF)~\cite{ZhangJ:2004} and the reliability ratio based weighted bit-flipping (RRWBF)~\cite{Guo:2004} ~\cite{Lee:2005_IRRWBF} algorithms try to narrow the gap between bit-flipping and sum-product algorithms by incorporating more information into the bit-flipping decision step.

In the LDPC decoding algorithms, four major steps are involved: initialization, check node update, variable node update, and decision.
The initialization step is executed only once.
The iteration is performed between the check node update step and the variable node update step.
The decision step is conducted after the stopping criterion is satisfied.
Before running the decoding algorithm, the magnitude of the received bits are converted to log-likelihood ratio (LLR) $r$.
\begin{equation}
r=\log \frac{\text{Prob}(y|x=1)}{\text{Prob}(y|x=-1)} \label{equ-LLR},
\end{equation}
where $x$'s are source bits and $y$'s are received bits.
For example, under Gaussian channel with noise variance $\sigma$ and BPSK modulation,
\begin{equation}
r=\frac{2}{\sigma^2}y \label{equ-LLR_AWGN_BPSK}.
\end{equation}
In the next few subsections, we will introduce them one by one.

\subsection{Sum-Product and log-SP Algorithm}
The first decoding algorithm is sum-product algorithm. The four steps are described below.
\begin{description}

\item[\underline{Initialization:}]
\begin{equation}
\alpha_{m,n}=r_n.
\end{equation}

\item[\underline{Check node update:}]
\begin{equation}
\beta_{m,n}=2\tanh^{-1}
\left(
\prod_{n' \in N(m)\setminus n}
\tanh
\left(\frac{\alpha_{m,n}}{2}
\right)
\right),
\end{equation}
where $(m,n) \in \{(i,j)|H_{i,j}=1\}$ and $N(m)$ is the set of variable nodes that are connected to the $m^{th}$ check node.\\

\item[\underline{Variable node update:}]
\begin{equation}
\alpha_{m,n}=r_n + \sum_{m' \in M(n)\setminus m} \beta_{m',n},
\label{equ-var}
\end{equation}
where $(m,n) \in \{(i,j)|H_{i,j}=1\}$ and $M(n)$ is the set of check nodes that are connected to the $n^{th}$ variable node.\\

\item[\underline{Decision:}]
\begin{equation}
\lambda_n = r_n + \sum_{m \in M(n)} \beta_{m,n}.
\end{equation}
\begin{equation}
\text{bit} = \left\{
\begin{array}{rl}
1 & \text{if } \lambda_n \geq 0\\
-1 & \text{if } \lambda_n < 0.
\end{array} \right.
\end{equation}\\

\end{description}

Log-SP is a variant of the sum-product algorithm. It can be derived by taking the logarithm of the sum-product algorithm. The four steps are:\\
\begin{description}

\item[\underline{Initialization:}]
\begin{equation}
\alpha_{m,n}=\text{sign}(r_n)\log\left(\tanh\left(\frac{|r_n|}{2}\right)\right).
\end{equation}

\item[\underline{Check node update:}]
\begin{equation}
\beta_{m,n}=-\log\left(\tanh\left(\frac{\alpha}{2}\right)\right)
\prod_{n' \in N(m)\setminus n}\text{sign}(\alpha_{m,n'}),
\end{equation}
where 
\begin{equation}
\alpha=\sum_{n' \in N(m)\setminus n} |\alpha_{m,n'}|,
\end{equation}
and $(m,n) \in \{(i,j)|H_{i,j}=1\}$.\\

\item[\underline{Variable node update:}]
\begin{equation}
\alpha_{m,n}=\text{sign}(\lambda_{m,n})\log\left(\tanh\left(\frac{|\lambda_{m,n}|}{2}\right)\right),
\end{equation}
where 
\begin{equation}
\lambda_{m,n}=r_n+\sum_{m' \in M(n)\setminus m}\beta_{m',n},
\end{equation}
and $(m,n) \in \{(i,j)|H_{i,j}=1\}$.\\

\item[\underline{Decision:}]
\begin{equation}
\lambda_n = r_n + \sum_{m \in M(n)} \beta_{m,n}.
\label{equ-decision}
\end{equation}
\begin{equation}
\text{bit} = \left\{
\begin{array}{rl}
1 & \text{if } \lambda_n \geq 0\\
-1 & \text{if } \lambda_n < 0.
\end{array} \right.
\end{equation}

\end{description}

\subsection{Min-Sum and Modified Min-Sum Algorithm}

The min-sum and the modified min-sum take the following formula.\\
\begin{description}

\item[\underline{Initialization:}]
\begin{equation}
\alpha_{m,n}=r_n.
\end{equation}

\item[\underline{Check node update:}]
\begin{equation}
\beta_{m,n}=\delta \cdot \prod_{n' \in N(m)\setminus n}\text{sign}(\alpha_{m,n'})
\min_{n' \in N(m)\setminus n}\vert\alpha_{m,n'}\vert,
\end{equation}
where $(m,n) \in \{(i,j)|H_{i,j}=1\}$.
If $\delta$ is 1, it is min-sum algorithm; otherwise, it is modified min-sum algorithm.\\

\item[\underline{Variable node update:}]
\begin{equation}
\alpha_{m,n}=r_n+\sum_{m' \in M(n)\setminus m}\beta_{m',n},
\end{equation}
where $(m,n) \in \{(i,j)|H_{i,j}=1\}$.\\

\item[\underline{Decision:}]
\begin{equation}
\lambda_n = r_n + \sum_{m \in M(n)} \beta_{m,n}.
\end{equation}
\begin{equation}
\text{bit} = \left\{
\begin{array}{rl}
1 & \text{if } \lambda_n \geq 0\\
-1 & \text{if } \lambda_n < 0.
\end{array} \right.
\end{equation}

\end{description}

\subsection{WBF and MWBF Algorithm}
BP-based algorithms yield excellent error-correcting capability, but their decoding complexity is also high.
BF-based LDPC code decoding algorithms, such as WBF and MWBF, are considered as a good trade-off between error-correcting performance and decoding complexity compared to BP-based decoding algorithms. 
Therefore, sometimes it is more practical to use bit-flipping decoding for energy-constrained mobile devices. 
The weighted bit-flipping and its modified version take the following four steps for decoding LDPC codes.\\

\begin{description}

\item[\underline{Initialization:}]
\begin{equation*}
\end{equation*}
Let $\bar{z}$ be the hard decision of $\bar{r}$ and
\begin{equation}
r^{min}_m=\min_{n \in N(m)}|r_n|.
\end{equation}

\item[\underline{Check node update:}]
\begin{equation}
s_m=\sum^{N}_{n=1} z_n H_{mn}.
\end{equation}

\item[\underline{Variable node update:}]
\begin{equation}
E_n=\sum_{m \in M(n)}(2s_m-1)r^{min}_{m}-\delta \cdot |r_n|.
\end{equation}
If $\delta=0$, it is WBF; otherwise, it is modified WBF.\\

\item[\underline{Decision:}]
\begin{equation*}
\end{equation*}
Flip the bit $z_n$ for
\begin{equation}
n=\text{arg}\max_n E_n.
\end{equation}

\end{description}

\subsection{RRWBF and IRRWBF Algorithm}

Guo and Hanzo showed that the reliability ratio based bit-flipping (RRWBF) algorithm performs best among existing bit-flipping-based algorithms ~\cite{Guo:2004}. The RRWBF algorithm is:\\
\begin{description}

\item[\underline{Initialization:}]
\begin{equation*}
\end{equation*}
Let $\bar{z}$ be the hard decision of $\bar{r}$ and
\begin{equation}
r^{max}_m=\max_{n \in N(m)}|r_n|,
\end{equation}
\begin{equation}
R_{mn}=\beta\frac{|r_n|}{|r^{max}_m|},
\end{equation}
where $\beta$ is a normalization factor.\\

\item[\underline{Check node update:}]
\begin{equation}
s_m=\sum^{N}_{n=1} z_n H_{mn}.
\end{equation}

\item[\underline{Variable node update:}]
\begin{equation}
E_n=\sum_{m \in M(n)}\frac{2s_m-1}{R_{mn}}.
\end{equation}

\item[\underline{Decision:}]
\begin{equation*}
\end{equation*}
Flip the bit $z_n$
\begin{equation}
n=\text{arg}\max_n E_n.
\end{equation}

\end{description}

The original RRWBF algorithm can be rewritten in a form to significantly reduce the decoding time.
This new algorithm is called implementation-efficient reliability ratio based bit-flipping algorithm (IRRWBF).
The details of the derivation can be found in Lee and Wolf's paper~\cite{Lee:2005_IRRWBF}.\\

\begin{description}

\item[\underline{Initialization:}]
\begin{equation*}
\end{equation*}
Let $\bar{z}$ be the hard decision of $\bar{r}$ and
\begin{equation}
T_m=\sum_{n \in N(m)}|r_n|.
\end{equation}

\item[\underline{Check node update:}]
\begin{equation}
s_m=\sum^{N}_{n=1} z_n H_{mn}.
\end{equation}

\item[\underline{Variable node update:}]
\begin{equation}
E_n=\frac{1}{|r_n|}\sum_{m \in N(m)}(2s_m-1)T_m.
\end{equation}

\item[\underline{Decision:}]
\begin{equation*}
\end{equation*}
Flip the bit $z_n$ for 
\begin{equation}
n=\text{arg}\max_n E_n.
\end{equation}

\end{description}

\begin{figure}[!t]
  \begin{center}
    \includegraphics[width=3.5in]{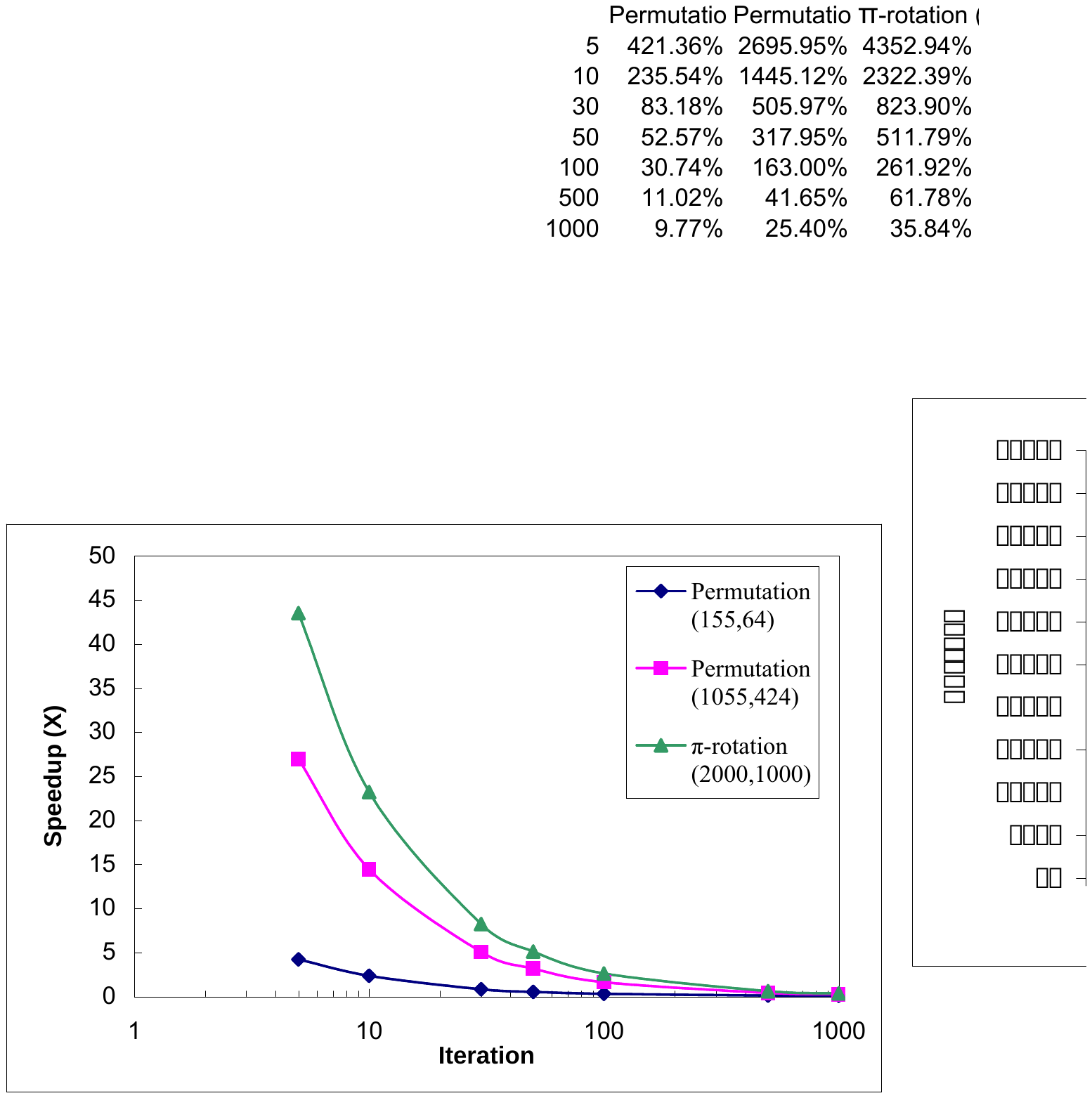}
  \end{center}
  \caption{Improvement of IRRWBF over RRWBF for some LDPC codes.}
  \label{fig-irrwbf}
\end{figure}

Fig.~\ref{fig-irrwbf} shows the runtime comparison of the original RRWBF and the IRRWBF algorithm~\cite{Lee:2005_IRRWBF}.
It shows that the IRRWBF runs much faster, especially when the iteration number is small, which is the case in most situations.

\subsection{Performance of LDPC Codes}

\begin{figure}[!t]
  \begin{center}
    \includegraphics[width=3.5in]{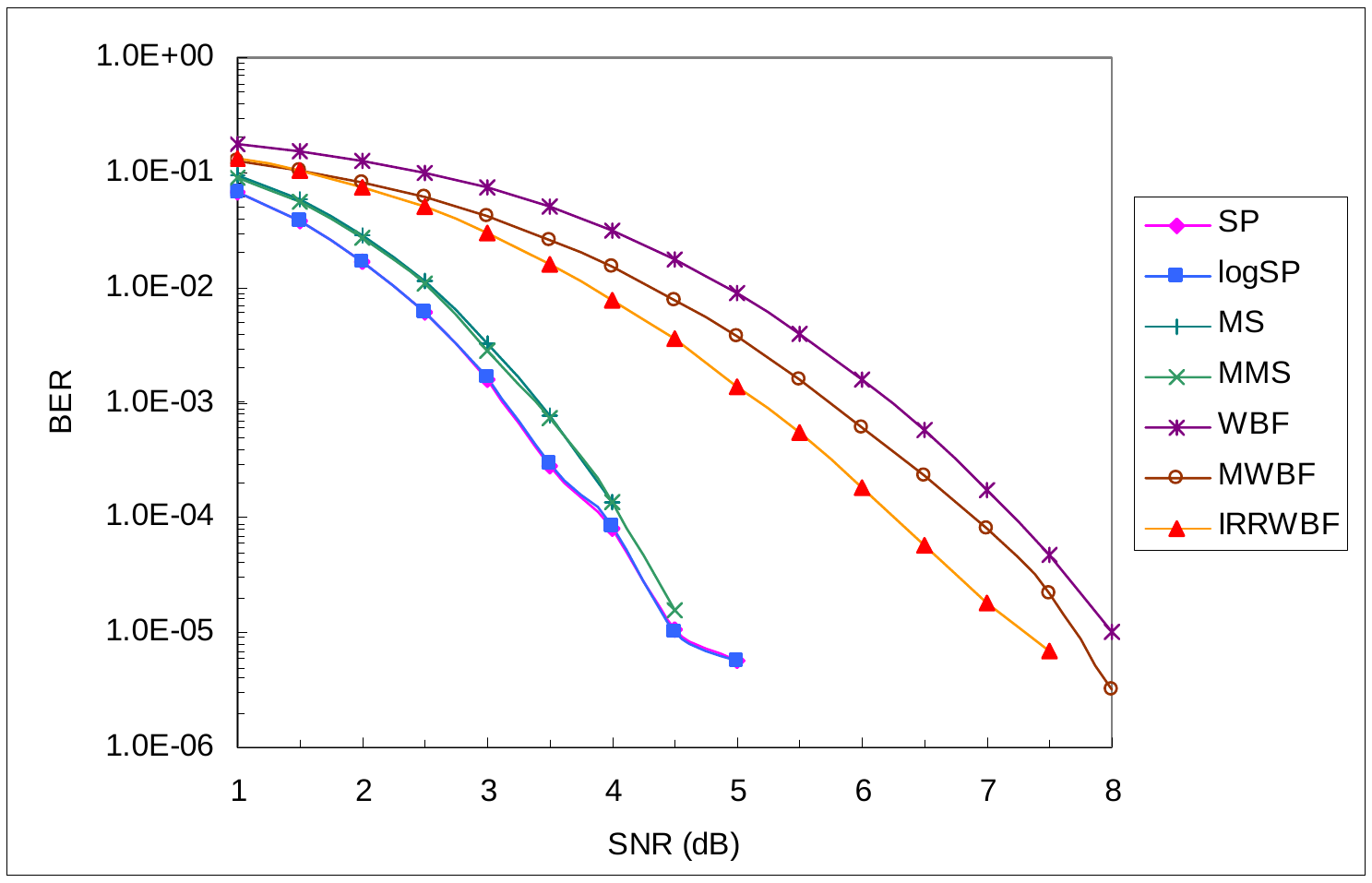}
  \end{center}
  \caption{Error rate of a (155, 64) LDPC code under AWGN channel.}
  \label{fig-ber}
\end{figure}

To compare the error-rate performance of different decoding algorithms, we chose the (155, 64) regular LDPC code constructed based on permutation matrices~\cite{Sridhara:2001} as an example. 
The stopping criterion for iteration is either $H_d^T = 0$ or 100 iterations is reached. 
The simulation results are shown in Fig.~\ref{fig-ber} for bit-error-rate (BER) performance~\cite{Lee:2005_Power}. 
AWGN channel is assumed in this setup. 
SP and log-SP algorithms give best error-correction for LDPC codes as expected. 
The performance of MS is close to the SP algo\-rithm, and the MMS algorithm performs slightly better than the MS decoding. 
IRRWBF algorithm performs best among BF-based algorithms, but it is still not as good as BP-based algorithms.

\subsection{Complexity of LDPC Codes}

When implementing an LDPC decoder, the complexity is important since it affects decoding rate and power consumption.
The complexity of different LDPC decoding algorithms can be estimated in the following way.
For a particular LDPC code, the connection between variable and check nodes is fixed.
Therefore, we can divide the problem of LDPC decoder implementation into two sub-problems: node processing and node connection. 
What mainly makes the complexity of decoding algorithms different is how messages are processed at nodes. 
Node connection, or the wiring, is almost the same for all algorithms.
In software implementation, in fact, there is no node connection.
As a result, to evaluate the difference in complexity, we can simply concentrate on comparing logic functions needed at check and variable nodes. 
Here we focus on the comparison of regular LDPC codes, i.e., codes that are generated by the parity-check matrix in which every row or column has the same number of nonzero entries. 
Let $W_c$ denote column weight (number of nonzero entries per column) and $W_r$ denote row weight (number of nonzero entries per row). 
Table \ref{table-complexity_variable} and Table \ref{table-complexity_check} show the usage of logic functions at variable and check nodes respectively for different decoding algorithms.
(Note that the decision steps of the sum-product, log-SP, min-sum, and MMS algorithm are the same, and the variable node steps of SP, MS, and MMS algorithm are the same.
In addition, the check node steps and the decision steps of the WBF, MWBF, RRWBF, and IRRWBF algorithm are the same.)
Multiplication and division operations usually are more complex than other operations, so it is expected that SP algorithm has higher complexity and will consume more power. 
Since most bit-flipping-based algorithms have simpler operations, their complexity and power consumption should be lower compared to SP, log-SP, MS and MMS algorithms.
Although only regular LDPC codes are considered here, the decoding complexity of irregular LDPC codes can be derived easily in the same way.

\begin{table}[!t]
  \caption{LDPC decoding complexity of variable nodes.}
  \begin{center}
    \includegraphics[width=3.5in]{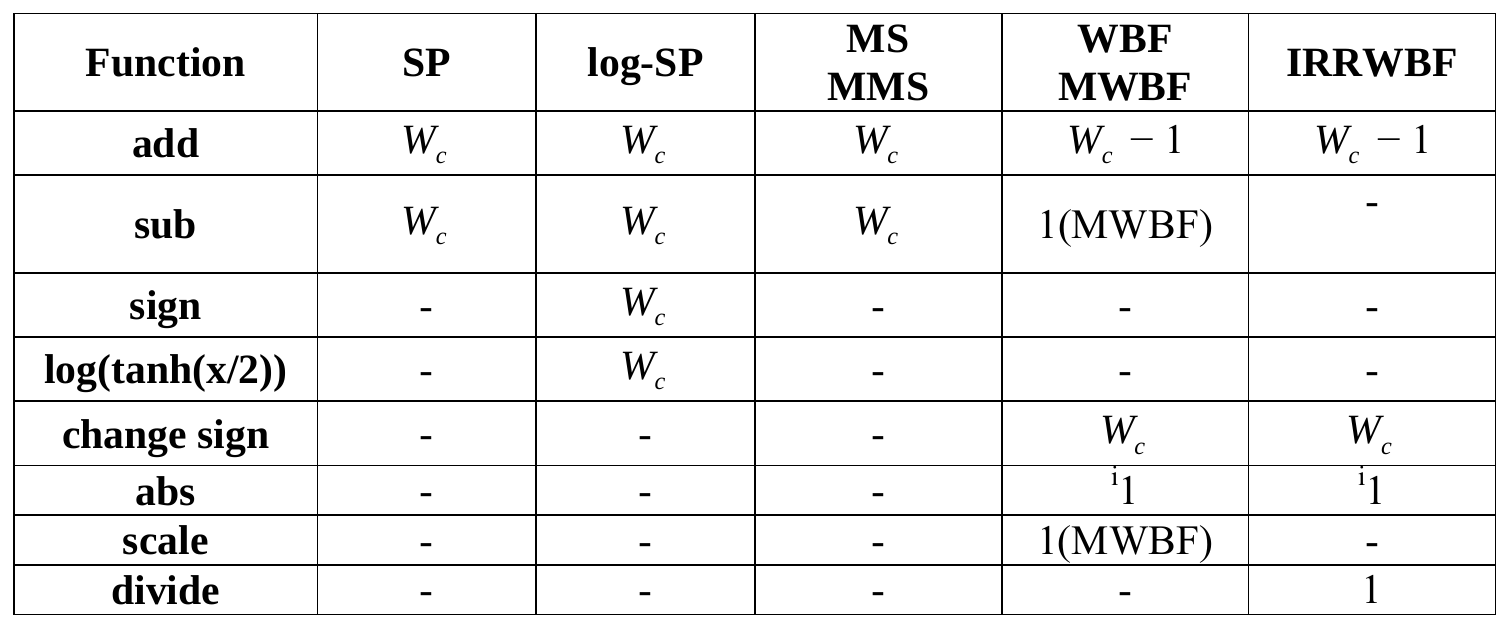}
  \end{center}
  \label{table-complexity_variable}
\end{table}

\begin{table}[!t]
  \caption{LDPC decoding complexity of check nodes.}
  \begin{center}
    \includegraphics[width=3.5in]{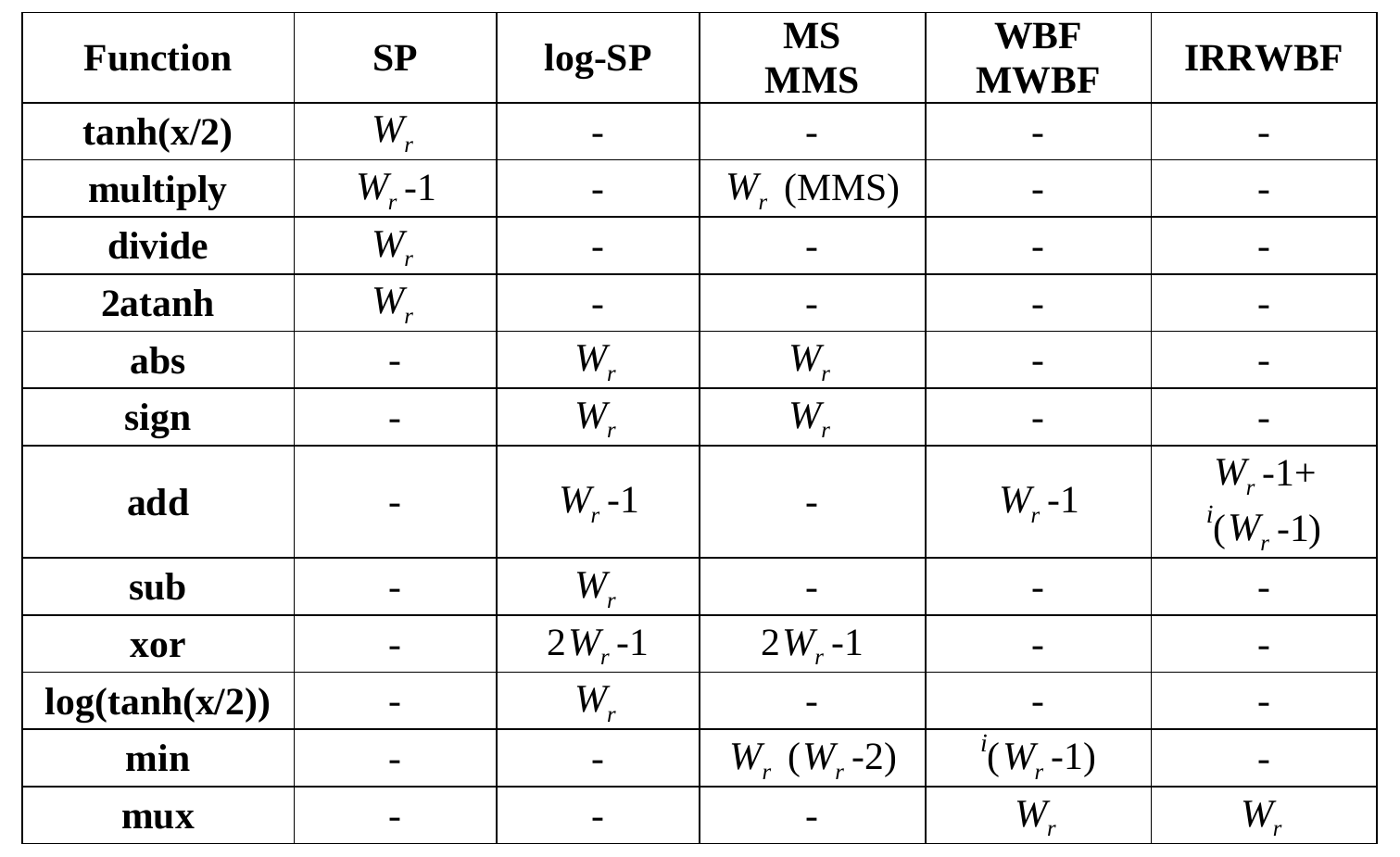}
  \end{center}
  \label{table-complexity_check}
\end{table}

\section{Processors as Software Radio Platform}\label{processor_platform}

\begin{figure}[!t]
  \begin{center}
    \includegraphics[width=2.5in]{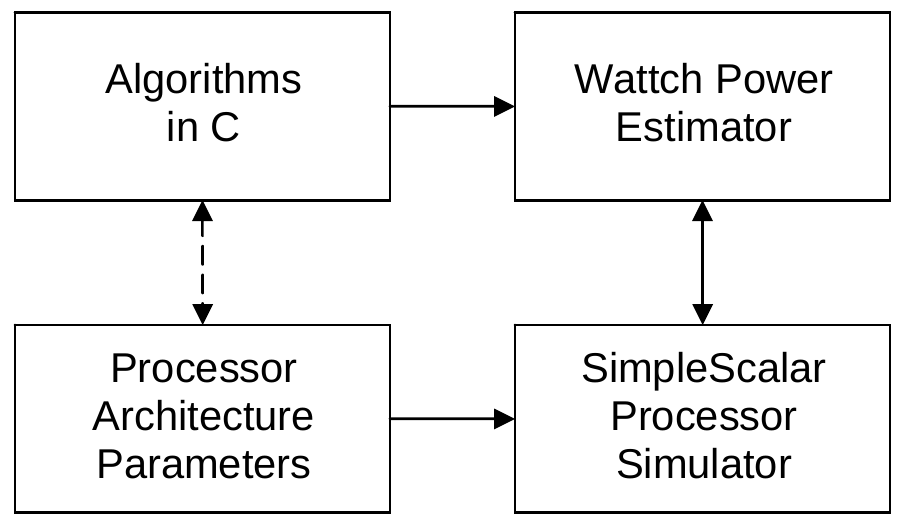}
  \end{center}
  \caption{Power estimation methodology.}
  \label{fig-method}
\end{figure}

Main concerns for the decoder implementation are performance, complexity, and power consumption. 
The lower the algorithm complexity, it is easier to finish before deadline and it also consumes less power.
Since processors consume more power than ASIC and FPGA, the complexity and power consumption are important issues.
The power consumption estimation for LDPC codes will be shown in part \ref{PowerConsumptionEstimation} and part \ref{ScaledPowerConsumptionEstimation}.

Optimization of algorithm performance on a processor is reflected in three aspects: algorithm itself, programming tricks, and the processor configuration.
The first step is to transform the algorithm into its most efficient form for the target platform.
An example is the RRWBF and the IRRWBF algorithm.
These two algorithms are basically the same, but their implementation complexity has significant difference.
For the programming part, commonly used programming techniques, such as in-lining and loop unrolling, should be applied regularly.
As to the processor configuration, parameters such as cache size and associativity, are critical to the performance and power consumption.
This will be shown in part \ref{secCache}.
Specialized operation or instruction sets for processor can also reduce runtime.

\begin{table}[!t]
  \caption{Processor configuration for power estimation.}
  \begin{center}
    \includegraphics[width=3.5in]{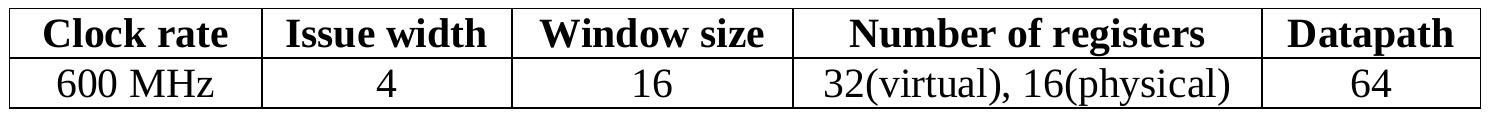}
  \end{center}
  \label{table-cpu_config}
\end{table}

\subsection{Power Consumption Estimation} \label{PowerConsumptionEstimation}

The proposed power estimation framework is sketched in Fig.~\ref{fig-method}. 
First of all, LDPC decoders are written in C. 
Instead of running LDPC decoders on a real GPP and then measuring their power consumption from the board, we use a power simulator. 
The advantages of using a simulator are fast and flexible.
By changing parameters, a simulator can model various GPP architectures. 
The simulator we use is Wattch~\cite{Brooks:2000}, which is a framework for analyzing and optimizing microprocessor power dissipation at the architecture-level. 
It claims to be 1000 times faster than the existing layout-level power estimation tools and yet maintains accuracy within 10\% of their estimates. 
Instead of using Wattch to simulate the power consumption of a computer architecture (as its original purpose), we feed our LDPC decoder into Wattch as a benchmark running on the specified GPP architecture with parameters shown in Table \ref{table-cpu_config}. 
Those parameters were set to reflect a real processor architecture. 
We compare the power dissipation based on the aggressive non-ideal (some fraction of power is still consumed when a functional unit is disabled) conditional clocking provided by Wattch. 
The decoders written in C are complied using GCC on a Solaris machine with level 2 optimization (-o2).

\begin{figure}[!t]
  \begin{center}
    \includegraphics[width=3.5in]{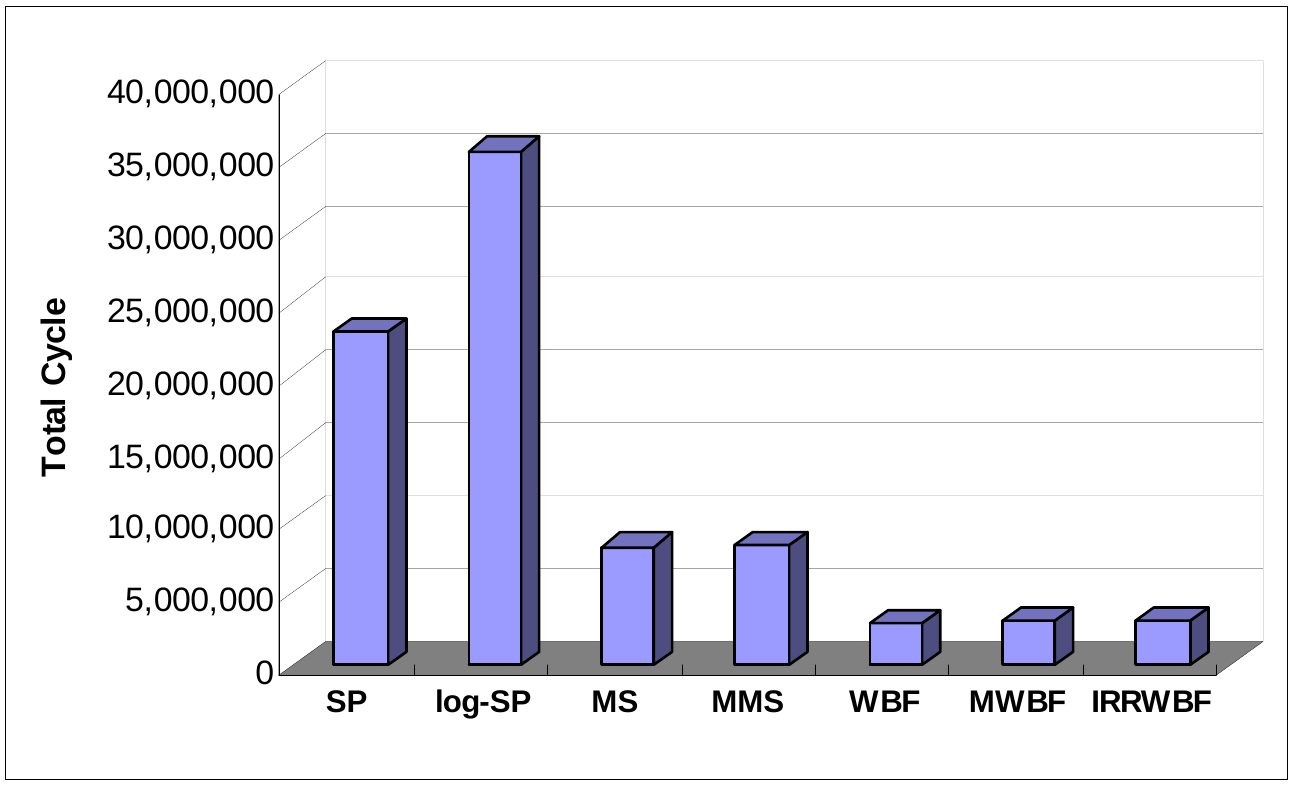}
  \end{center}
  \caption{Total number of cycles for different decoding algorithms.}
	\label{fig-total_cycle}
\end{figure}

\begin{figure}[!t]
  \begin{center}
    \includegraphics[width=3.5in]{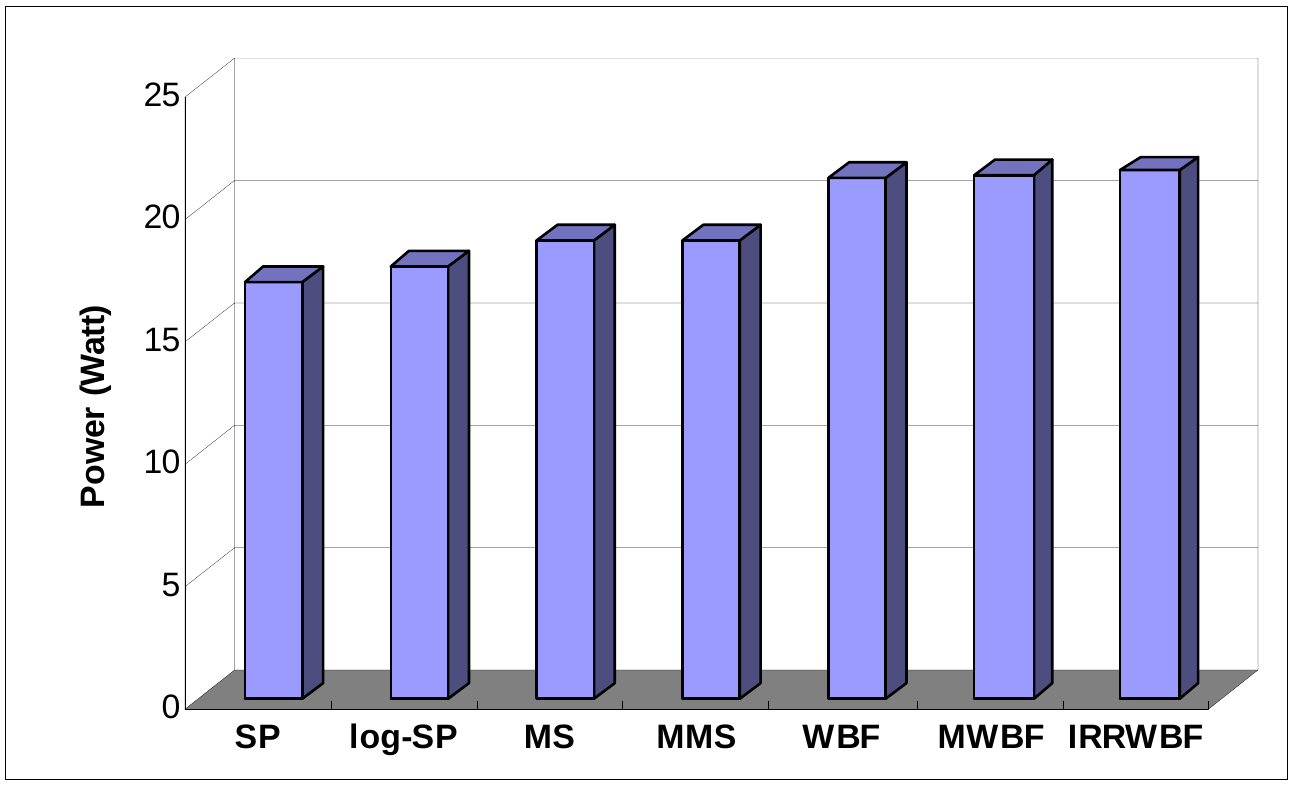}
  \end{center}
  \caption{Power dissipation for different decoding algorithms.}
	\label{fig-power}
\end{figure}

\begin{figure}[!t]
  \begin{center}
    \includegraphics[width=3.5in]{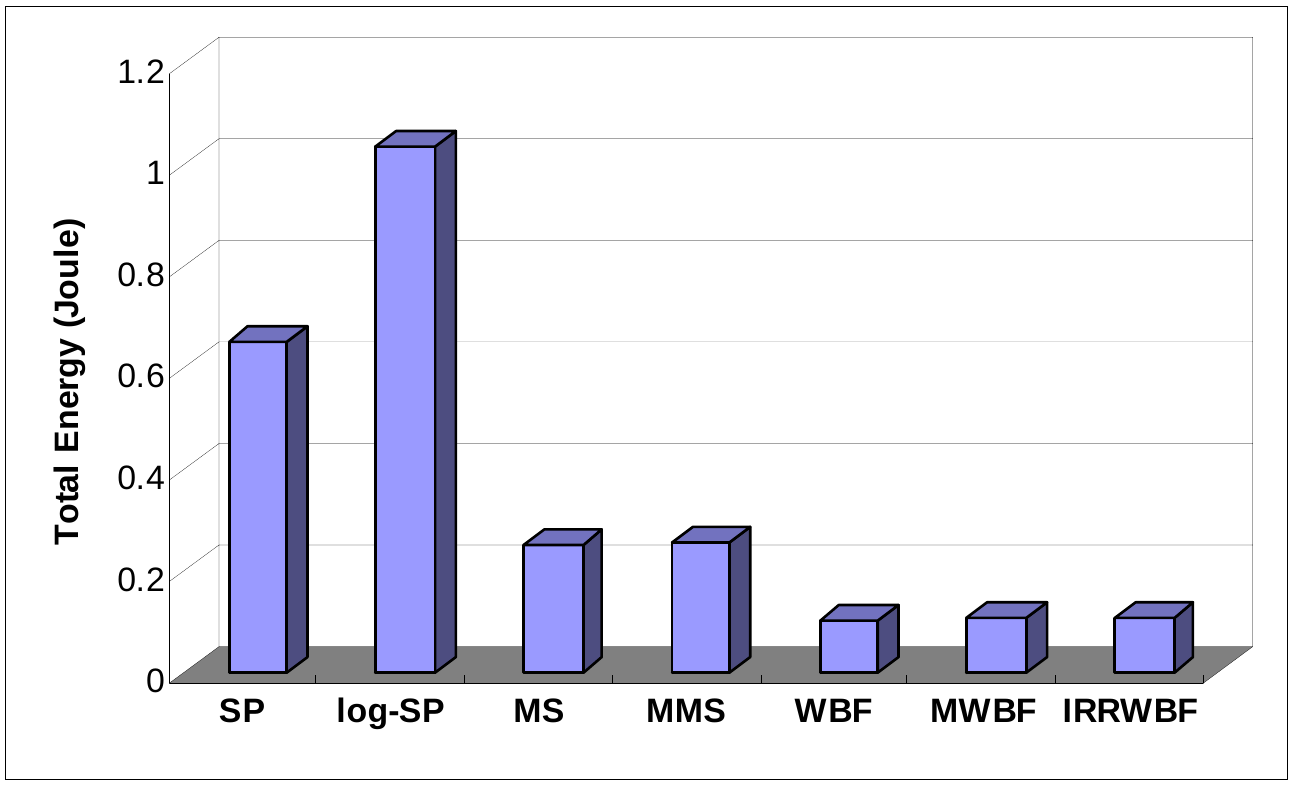}
  \end{center}
  \caption{Energy consumption for different decoding algorithms.}
	\label{fig-energy}
\end{figure}

\begin{figure}[!t]
  \begin{center}
    \includegraphics[width=3.5in]{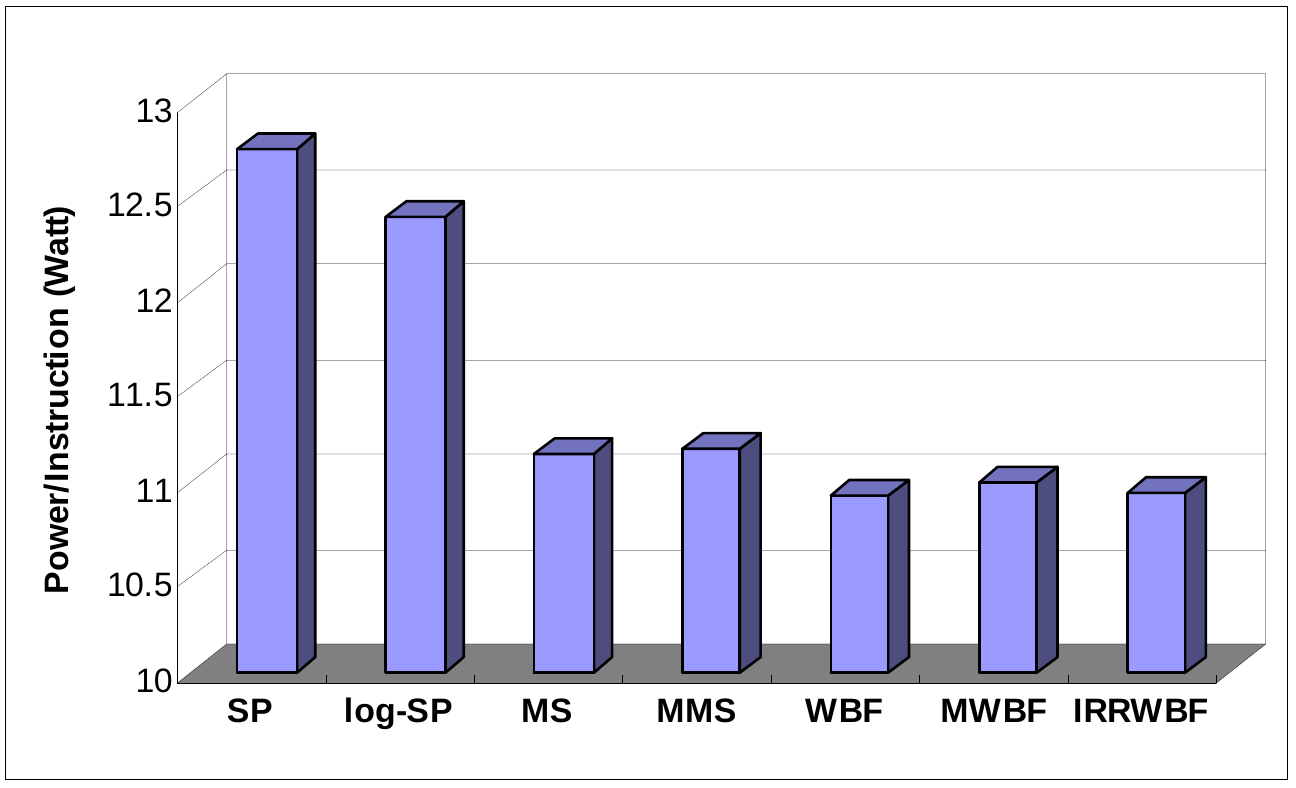}
  \end{center}
  \caption{Power per instruction for different decoding algorithms.}
	\label{fig-power_instruction}
\end{figure}

\begin{figure}[!t]
  \begin{center}
    \includegraphics[width=3.5in]{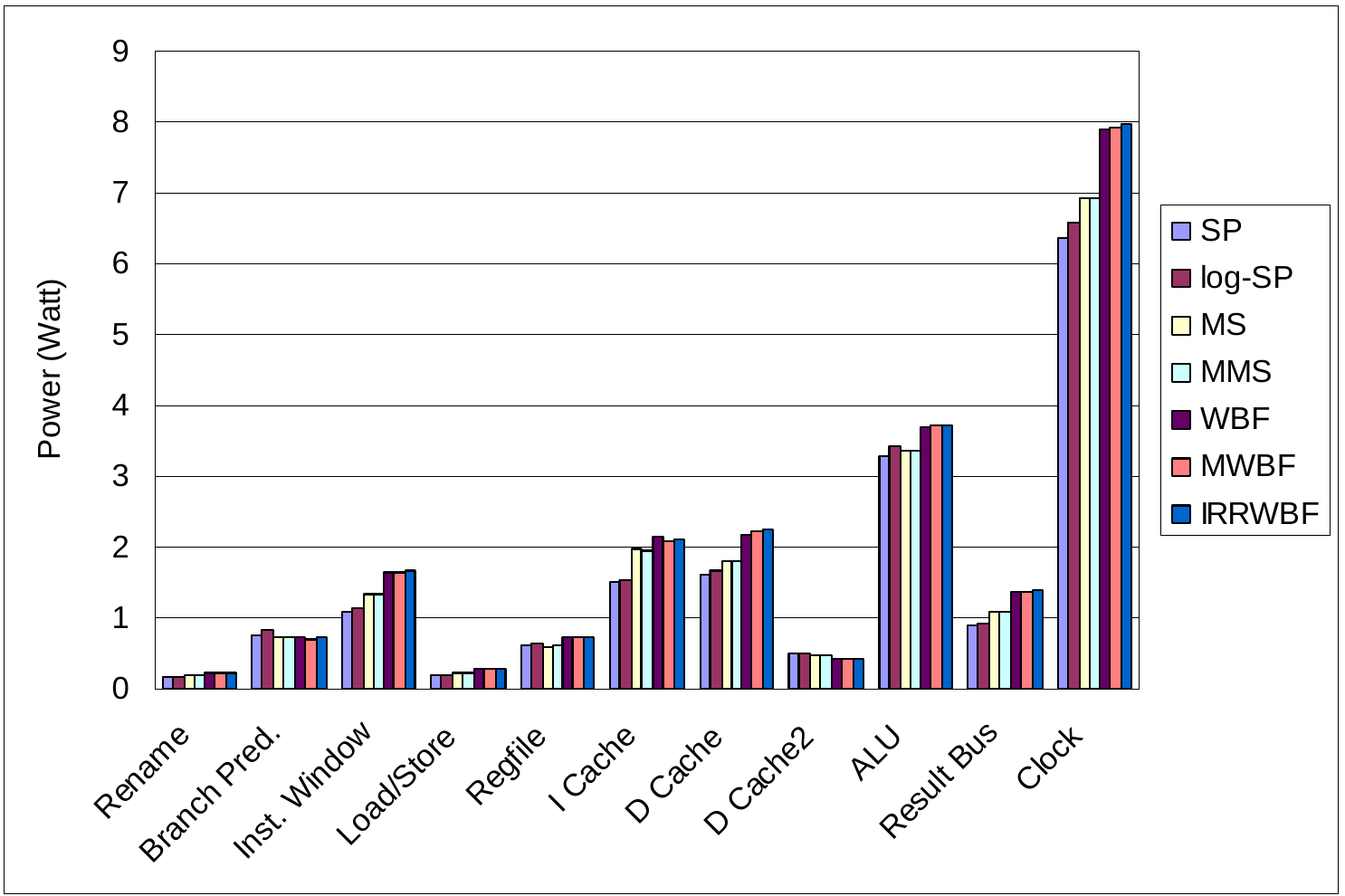}
  \end{center}
  \caption{Breakdown of power dissipation for different decoding algorithms.}
	\label{fig-power_breakdown}
\end{figure}

\begin{figure}[!t]
  \begin{center}
    \includegraphics[width=3.5in]{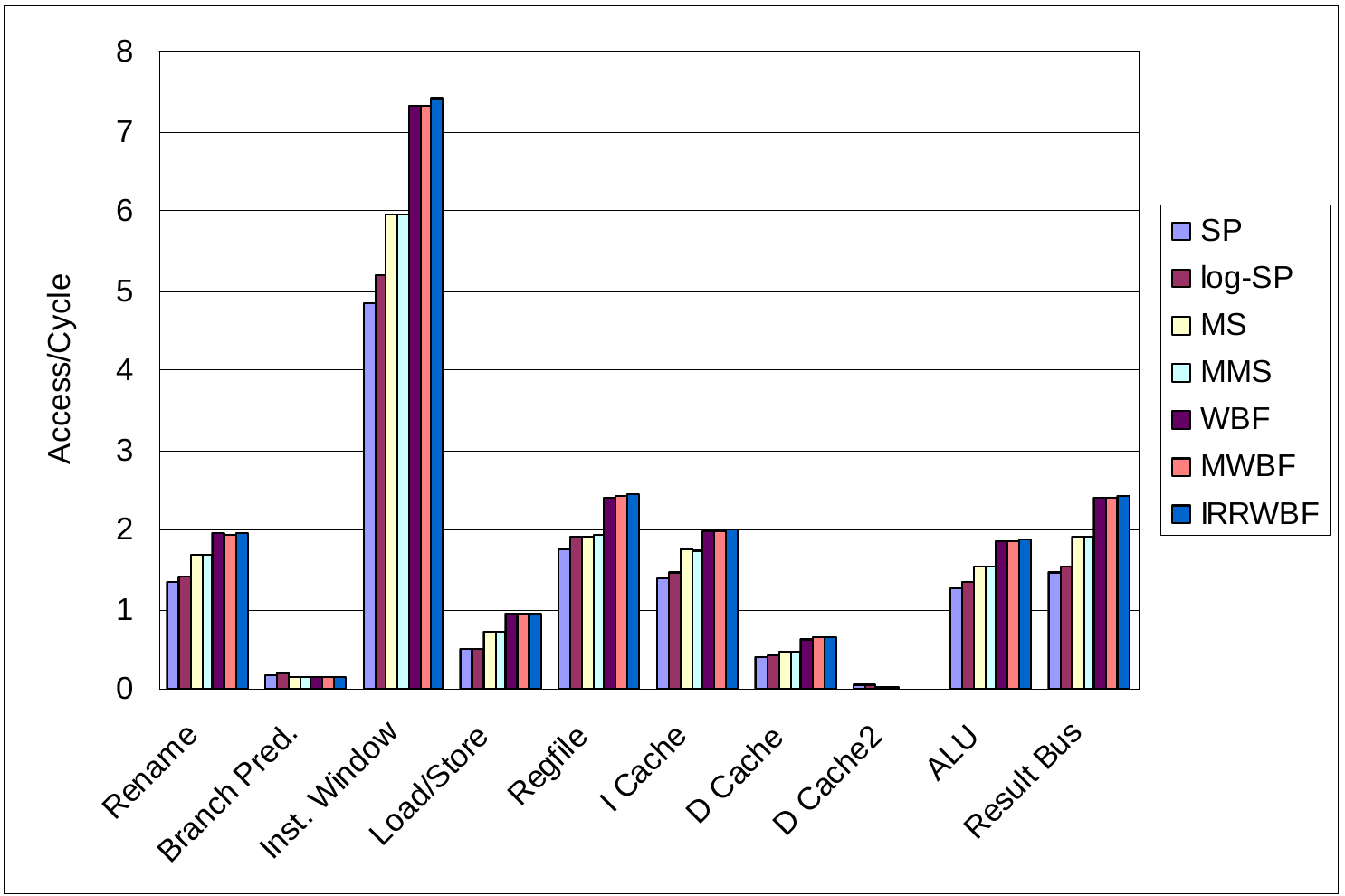}
  \end{center}
  \caption{Breakdown of number of access for different decoding algorithms.}
	\label{fig-access_breakdown}
\end{figure}

The detailed power consumption statistics for different decoding algorithms are extracted from the Wattch reports.
During simulation, we set the iteration number to 100 for all algorithms without using the early stopping criterion. 
This gives us an estimate of power consumption per iteration when we divide this value by the number of iteration, which is 100 in our setting.
The simulation results are shown in Fig.~\ref{fig-total_cycle}, Fig.~\ref{fig-power}, Fig.~\ref{fig-energy}, and Fig.~\ref{fig-power_instruction}.
Fig.~\ref{fig-total_cycle} shows total GPP cycles, Fig.~\ref{fig-power} shows dissipated power, Fig.~\ref{fig-energy} shows total energy consumption, and Fig.~\ref{fig-power_instruction} shows power per instruction for different decoding algorithms, all with 100 iterations.

An interesting result is: although log-SP algorithm is popular than SP in hardware implementation~\cite{ZhangT:2004}, it is inefficient in terms of energy consumption for decoders implemented in software. 
The reason is that in hardware implementations, $log(tanh(x/2))$ can be easily implemented using a lookup table (since messages are quantized), so it would be advantageous to have the extra $log(tanh(x/2))$ functions at the variable nodes and the extra $log(x)$ functions at the check nodes to trade for multiplications and divisions. 
The software implementation of $log$ and $tanh$ functions, however, takes much longer time than multiplication operations, so log-SP consumes more power. 
Writing a look-up-tables (LUT) in the code is possible, but it will sacrifice the error-correction performance due to quantization error unless the LUT is large. 

MS and MMS algorithms remove the need of $tanh(x/2)$ and some adders or multipliers, resulting in a huge drop in both GPP execution cycle count and total energy consumption.
The energy consumption of bit-flipping algorithms is only about one-sixth of SP and one half of MS algorithms. 
Results from Fig.~\ref{fig-total_cycle} also suggest that if the decoding is delay-constrained, bit-flipping-based algorithms are better candidates than BP-based algorithms.

While energy consumption is related to battery lifetime, power level is relevant to heating up the circuits. 
Fig.~\ref{fig-power} shows that IRRWBF has the highest power dissipation, so it is undesirable if chip cooling is an issue. 
The power consumption of bit-flipping-based algorithms is larger than that of MS and SP algorithms. 
To further investigate this, power dissipation in each GPP hardware unit is compared in Fig.~\ref{fig-power_breakdown}, and Fig.~\ref{fig-access_breakdown} shows the number of access to different processor functional units. 
All these decoding algorithms dissipate almost the same power in rename, branch prediction, load/store queue, register files, and level-2 data cache units. 
However, bit-flipping-based algorithms dissipate more power than other algorithms in instruction window, instruction cache, level-1 data cache, ALU, result bus, and clock units. 
Their cache miss rate is lower since level-2 data cache is accessed less and level-1 data cache is accessed more. 
This infers that pipeline stall resulted from cache miss penalty happens less.
According to the power model, units are disabled when not in use, so their clock power consumption would be less (but nonzero due to leakage) during the pipeline stall.
That explains why the power dissipation of WBF, MWBF, and IRRWBF is higher than SP and log-SP algorithms.

\subsection{Power Consumption Estimation with Stopping Criterion} \label{ScaledPowerConsumptionEstimation}

Different decoding algorithms require different number of iterations to successfully decode a codeword. 
Even for a particular decoding algorithm, iteration number varies depending on the received bits. 
This makes the energy consumption simulation for decoders with early stopping mechanism very time-consuming if we try to feed thousands of blocks into the Wattch simulator to make the results statistically right. 
To give an idea of how much time it would cost, let us consider the following case. 
It takes 80 seconds to simulate one block with 10 iterations for SP algorithm. 
To get a statistically accurate result, we need to simulate at least 1000 to 10000 blocks, and that will take one to ten days to finish only one simulation for one SNR using one CPU unit. 
Then it will take over one year to finish all the energy consumption estimation presented here. 
Using a server with multiple CPUs runs faster but is more expensive. 
That motivates the need of an efficient method to estimate energy consumption when the early stopping mechanism is applied.

\begin{figure}[!t]
  \begin{center}
    \includegraphics[width=3.5in]{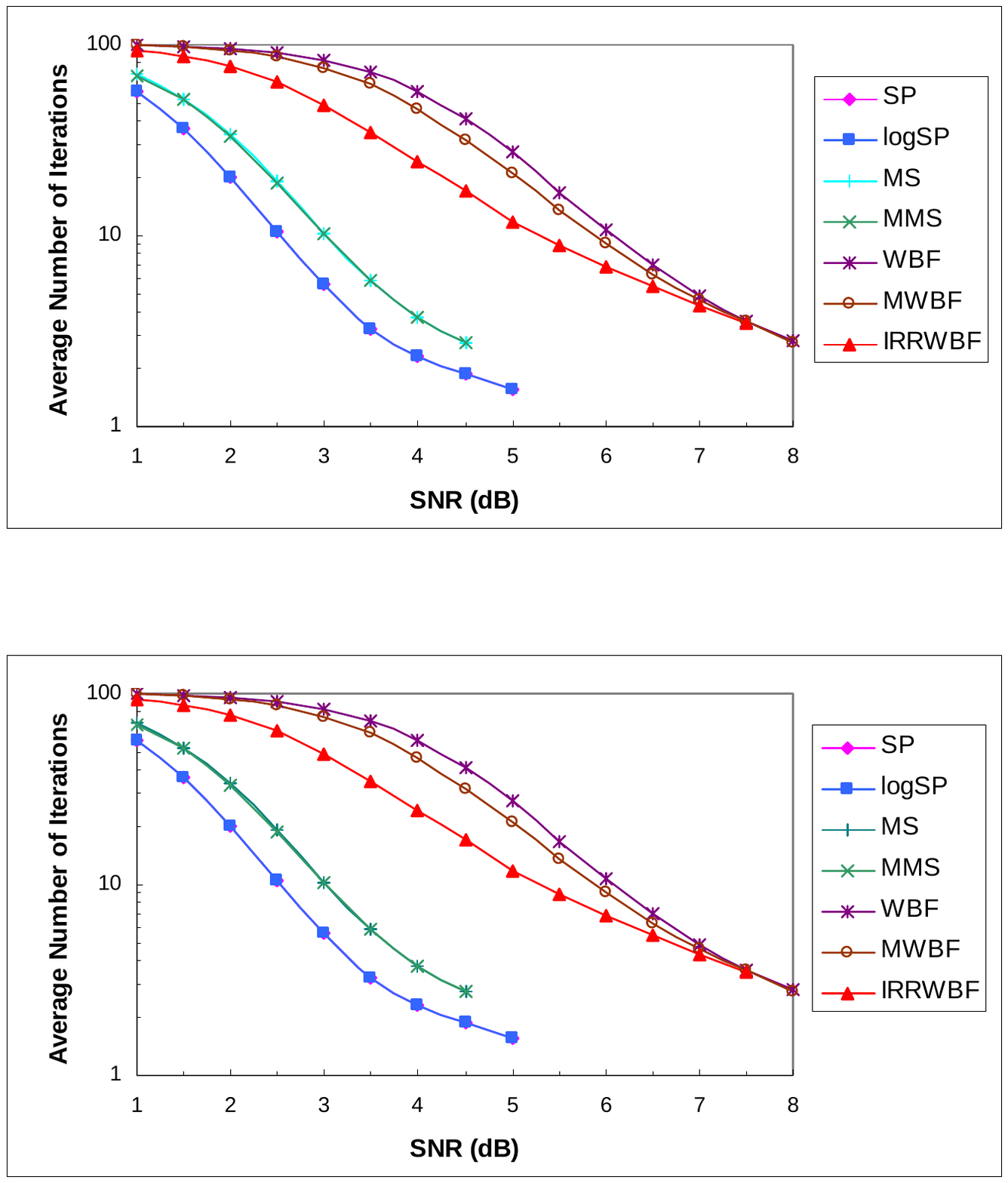}
  \end{center}
  \caption{Average number of iterations required for decoding.}
  \label{fig-average_iteration}
\end{figure}

The main idea of the proposed method is to gather the statistics information through simulating the average number of iterations needed to decode a LDPC code under different SNR for different algorithms. 
The methodology is summarized below, and the (155, 64) regular LDPC code is used as an example.

\begin{figure}[!t]
  \begin{center}
    \includegraphics[width=3.5in]{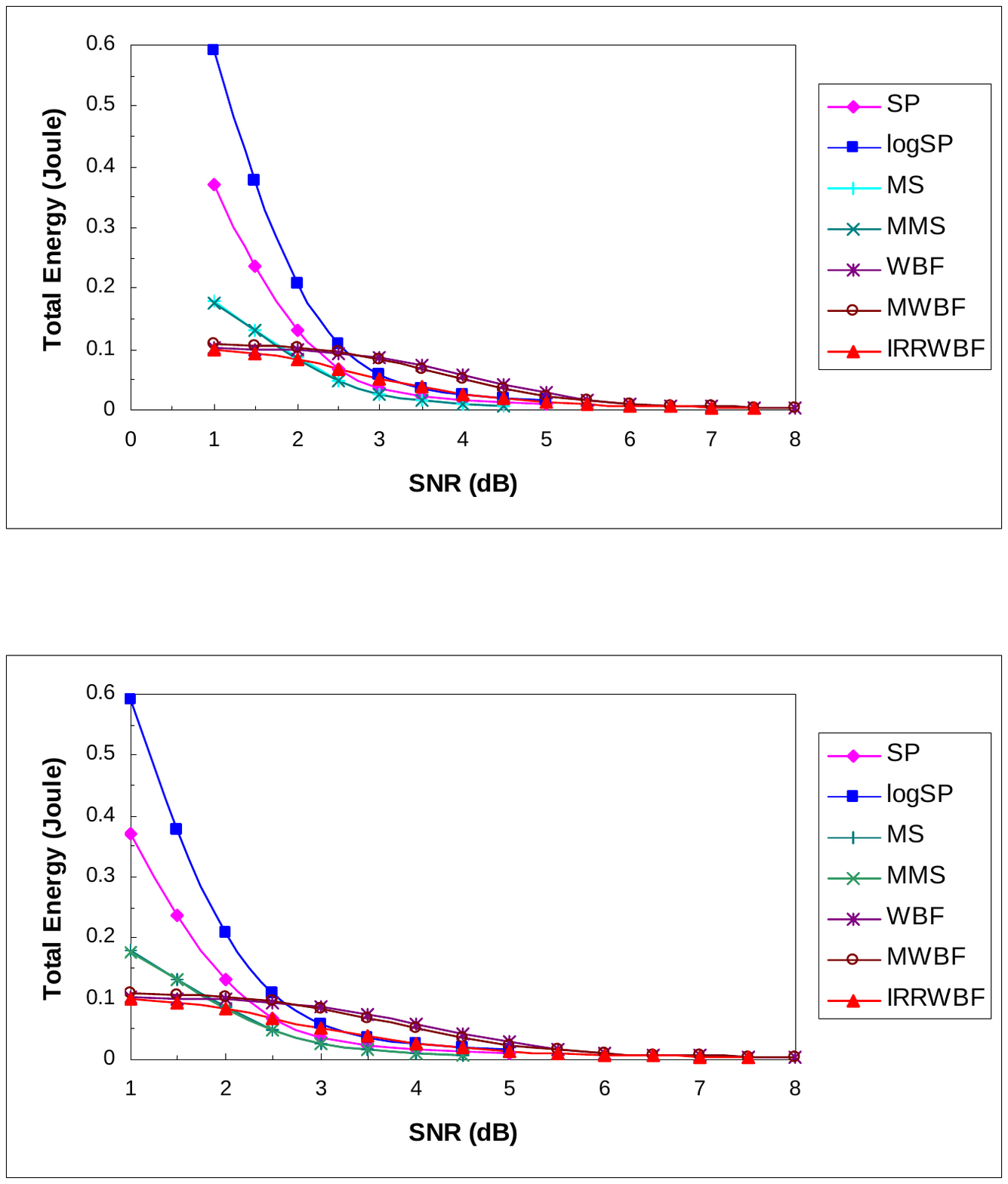}
  \end{center}
  \caption{Energy consumption for different decoding algorithms.}
  \label{fig-scaled_energy}
\end{figure}

\begin{itemize}
\item	Step 1: Simulate the energy consumption of different decoding algorithms using fixed number of iterations (say 100 iterations) and then normalize the results.
\item	Step 2: Simulate to get the average number of iterations under different SNR for different algorithms (Fig.~\ref{fig-average_iteration}). This step gets statistical results, and can be done in C or Matlab.
\item	Step 3: Scale the power consumption by the iteration number for different algorithms under different SNR to get energy consumption versus SNR plot (Fig.~\ref{fig-scaled_energy}).\\
\end{itemize}

The resulting plot in Fig.~\ref{fig-scaled_energy} shows the total energy consumption when applying early stopping criterion.
It basically says: for higher SNR ($>$ 2dB), MS and MMS are the best algorithms in terms of power consumption. When SNR is below 2dB, IRRWBF consumes less power.

\subsection{Decoding Data Rate}

Other than power consumption, decoding data rate is an important metric of the algorithm performance. 
To calculate the decoding data rate, we use the following formula: 

\begin{equation}
\begin{array}{c}
\text{Decoding rate} = \text{Processor frequency} / \\
(\text{Cycle per iteration} * \text{Iteration number}) \\ 
* \text{Bit length} * (1-\text{BER})
\end{array}
\end{equation}

\begin{figure}[!t]
  \begin{center}
    \includegraphics[width=3.5in]{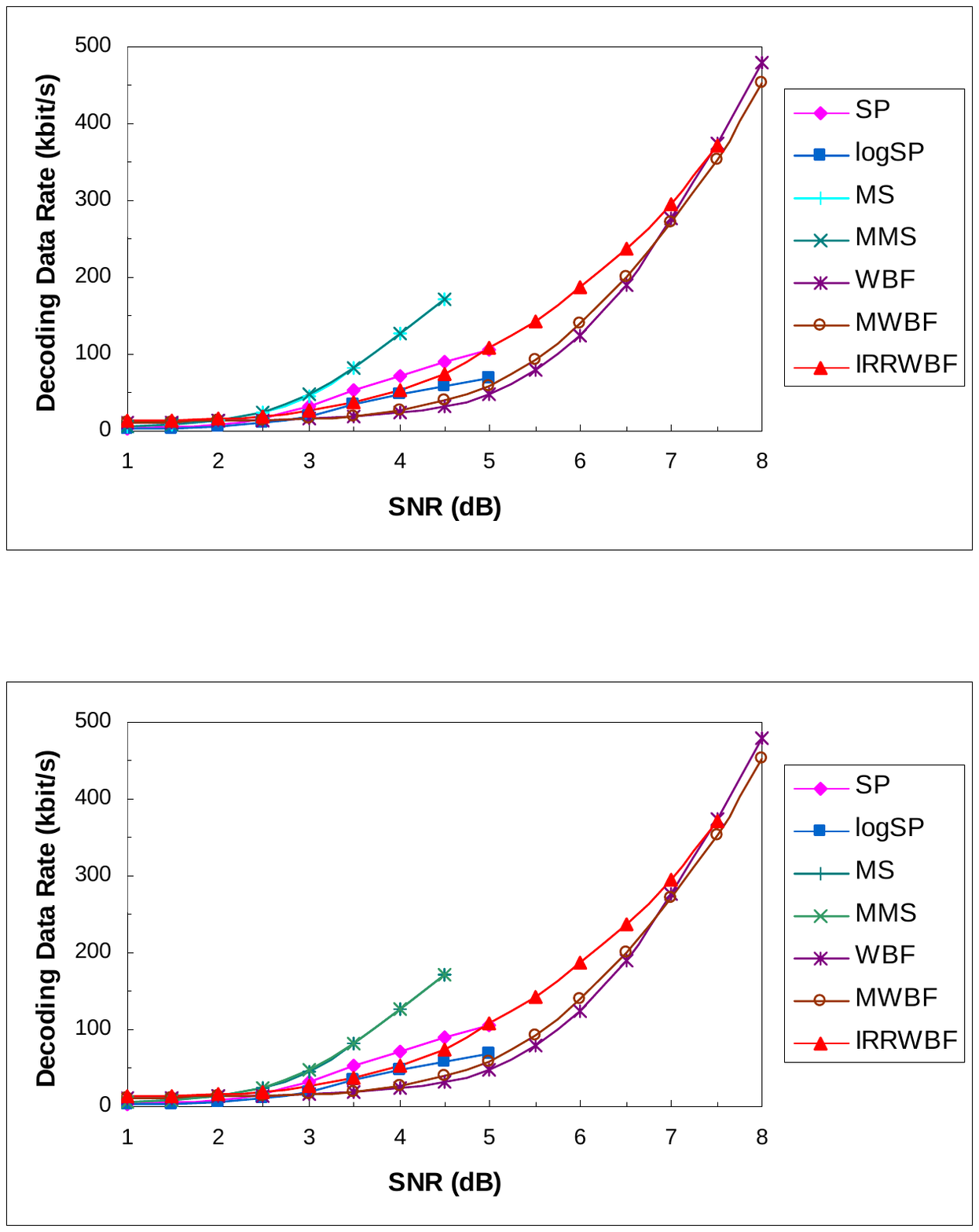}
  \end{center}
  \caption{Decoding data rate using different decoding algorithms.}
  \label{fig-decoding_data_rate}
\end{figure}

The decoding data rate depends on how fast a processor runs (processor frequency), the algorithm complexity, and the error-correcting performance (bit-error-rate).
Cycle per iteration is actually the normalization of Fig.~\ref{fig-total_cycle}.
Iteration number is shown in Fig.~\ref{fig-average_iteration} and BER is shown in Fig.~\ref{fig-ber}.
If the processor frequency is set to 600 MHz and the bit length of the LDPC code is 64, we get the simulation results presented in Fig.~\ref{fig-decoding_data_rate}.
It shows MS/MMS has higher decoding data rate, which means it is more efficient to choose an algorithm with moderate complexity and good enough performance, instead of the best performance.

\subsection{Cache} \label{secCache}

In order to understand how cache parameters play the role in performance of the software radio implementation, we need to turn to real processors.
Tensilica provides customized configurable processor, which allows us to choose the cache size, cache associativity, floating point unit, and design instruction set extension~\cite{tensilica}.
By using Tensilica's development tools for Xtensa processors~\cite{xtensa}, we are able to perform experiments on how the cache affects the performance of a specific implementation and then choose the best configuration.

\begin{figure}[!t]
  \begin{center}
    \includegraphics[width=3.5in]{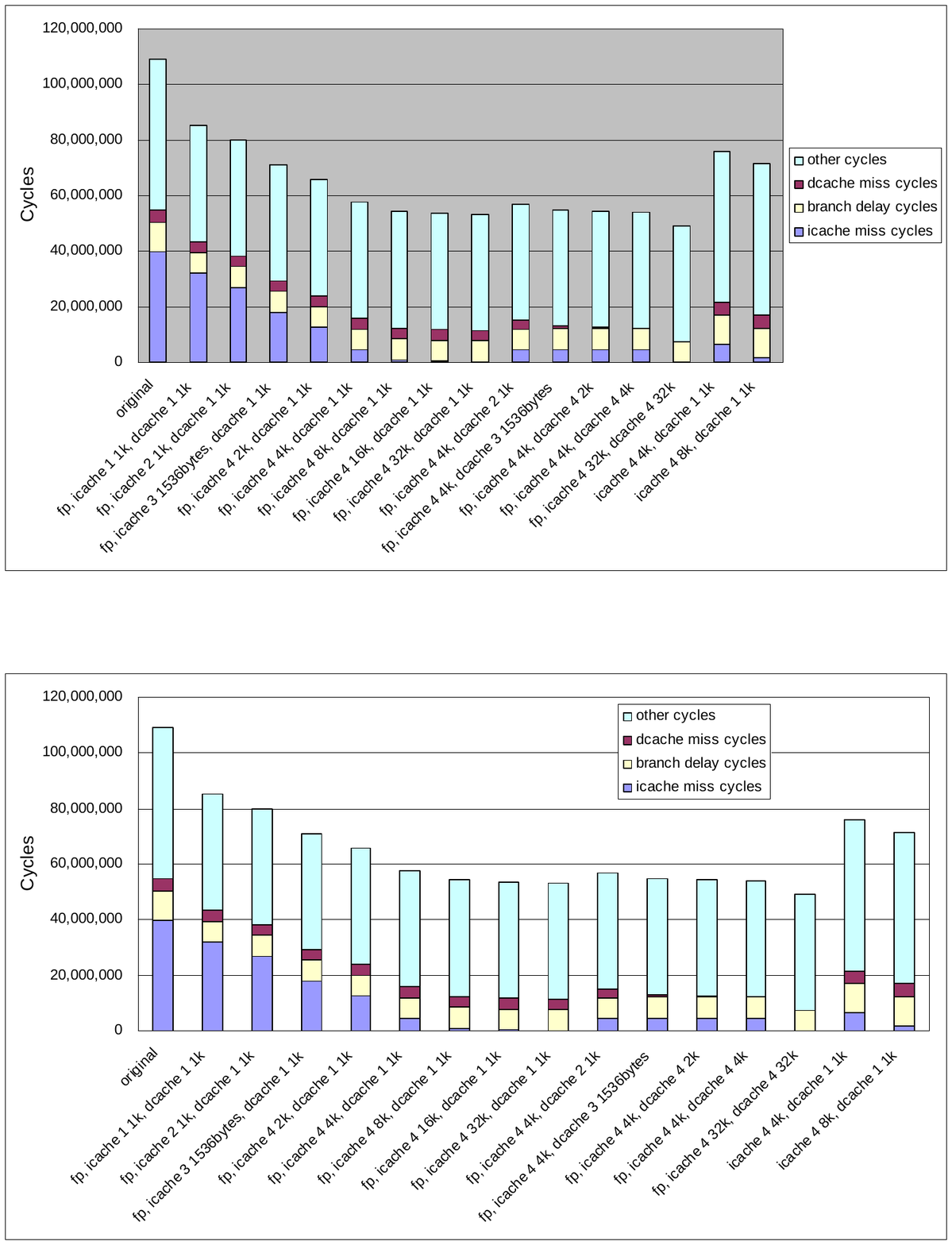}
  \end{center}
  \caption{Number of cache cycles for different cache configurations.}
  \label{fig-cache_cycle}
\end{figure}

\begin{figure}[!t]
  \begin{center}
    \includegraphics[width=3.5in]{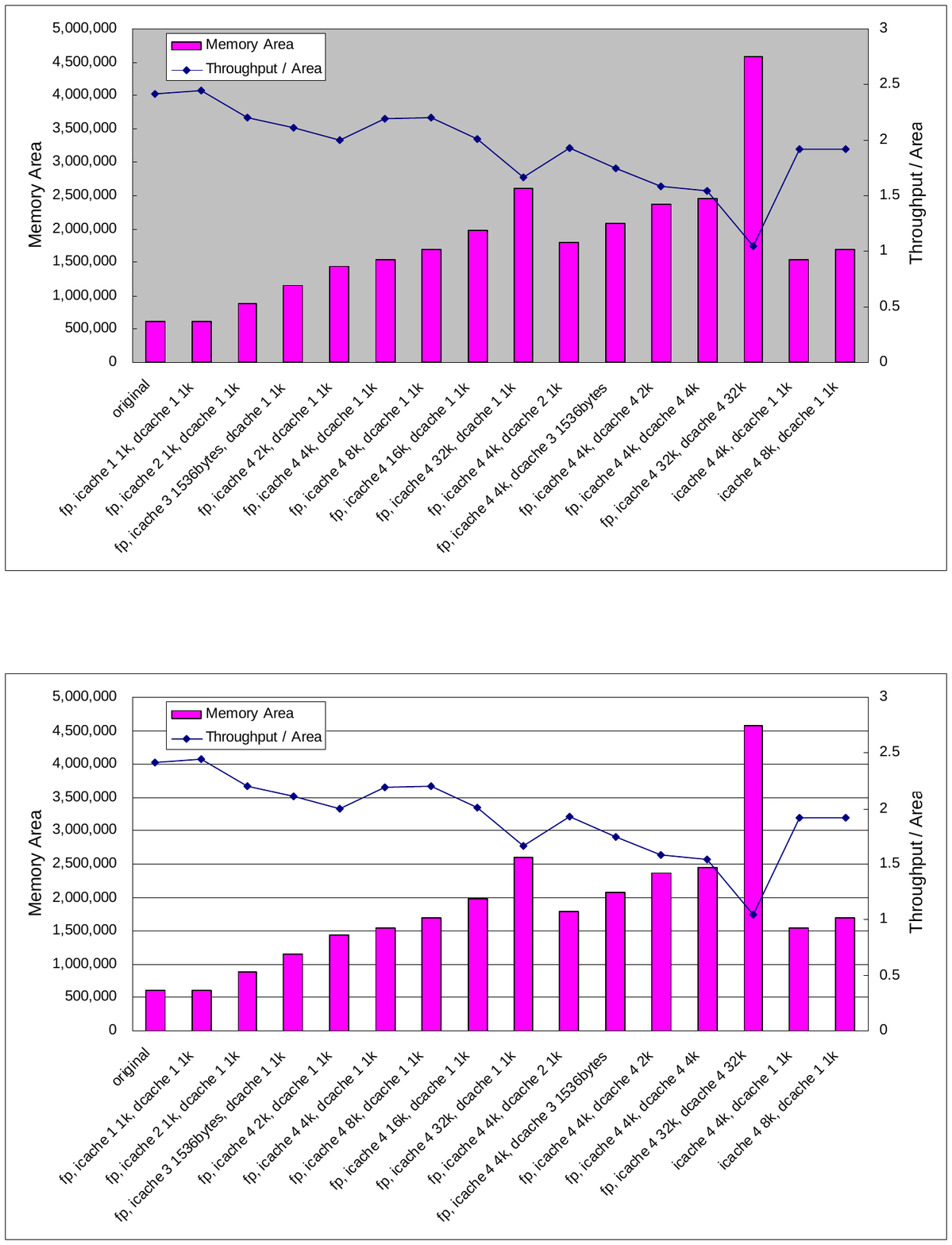}
  \end{center}
  \caption{Cache area for different configurations.}
  \label{fig-cache_area}
\end{figure}

We choose sum-product algorithm as the benchmark to obtain cache cycles (Fig.~\ref{fig-cache_cycle}) and cache area (Fig.~\ref{fig-cache_area}) for different configurations.
In these two figures, \emph{icache} means instruction cache, \emph{dcache} means data cache and \emph{fp} means floating point unit.
``fp, icache 4 4k, dcache 1 1k'' thus means the floating point unit is activated, the instruction cache size is 4k bytes with 4-way associativity, and the data cache size is 1k bytes with no associativity.
In general, the larger the cache size or the larger the cache associativity, the lower the cache miss rate is.
By increasing the cache size, the cache miss rate can be lowered and the total cycle count can be reduced.
Fig.~\ref{fig-cache_cycle} shows data cache miss is not a significant factor compared to instruction cache miss so minimizing the instruction cache cycle is more critical.
The instruction cache cycle count can become very small when we keep increasing the cache size and associativity, but we have to pay the price of higher power consumption and larger cache area (See Fig.~\ref{fig-cache_area}).
As a result, choosing 4k bytes of instruction cache is good enough for the performance.
Choosing 8k bytes of instruction cache is wasted and unnecessary for this algorithm on this platform.

\section{Power Control Schemes}\label{power_control}

The emergence of software radio concept leads people to consider implementing decoders in software, and therefore, the possibility of dynamically switching between different decoding algorithms to adapt to the environments and channels. This is termed \emph{algorithm diversity} in some literatures~\cite{Atluri:2004}~\cite{Karasawa:2000}. 
The concept is to select the algorithm which works most efficiently under each circumstance. 
Efficiency could mean having the best error-correcting ability \emph{with the constraints in delay and power consumption}. 
Therefore, if the battery on a mobile device is dying, it would be pointless to use power consuming decoding algorithms even they have good error-correcting performance. 
Under this situation, the decoder can trade performance for power consumption. 
In addition, different standards specify different delay constraints. 
If using a powerful but complex decoding algorithm cannot meet the delay requirement, switching to a simpler decoding algorithm with less error-correcting capability  is desired. 
Existing decoding algorithms for LDPC codes have trade-offs between bit-error-rate, delay, and power consumption, so it would be a good example for algorithm diversity.
In this section, we will describe two power control schemes: SNR-based algorithm diversity and joint transmit power and receiver energy management.

\subsection{SNR-Based Algorithm Diversity} \label{SNRalgorithmDiversity}

Consider a mobile terminal (such as a cell phone) with software radio capability.
When it moves around, the channel between the transmitter and the receiver is time-varying.
Suppose the signal-to-noise ratio is detectable at the receiver, we propose a simple algorithm diversity control scheme.
The first step is to define an SNR-related performance metric, such as power consumption or decoding rate.
This proposed algorithm is to choose the algorithm with the best performance under current channel condition in terms of SNR.
To illustrate this scheme, we utilize the results presented in Fig.~\ref{fig-scaled_energy} and Fig.~\ref{fig-decoding_data_rate}.
If we consider energy consumption as the performance metric as in Fig.~\ref{fig-scaled_energy}, this control scheme chooses IRRWBF if SNR is between 1 dB and 2 dB, chooses modified min-sum if SNR is between 2 dB and 3 dB, chooses min-sum algorithm if SNR is between 3 dB and 4 dB, and again chooses MMS if SNR is between 4 dB and 4.5 dB.
If we consider the decoding data rate as the metric (Fig.~\ref{fig-decoding_data_rate}), this scheme chooses IRRWBF when SNR falls between 1 dB and 2 dB, chooses MMS if SNR falls between 2 dB and 3.5 dB, and then chooses min-sum if SNR is between 3.5 dB and 4.5 dB.
In this example, the high data rate and low energy algorithm at each SNR happens to be almost the same.
If there is a conflict but one wants to satisfy both criterion, a cost function must be defined to resolve this problem. 
Since the results of Fig.~\ref{fig-scaled_energy} and Fig.~\ref{fig-decoding_data_rate} may be different when implemented on a platform with different processor parameters, these figures must be reproduced under different scenarios.



\subsection{Joint Transmit Power and Receiver Energy Management}

The control scheme described in part \ref{SNRalgorithmDiversity} only considers power consumption at the receiver. 
Here we propose a joint transmit power and receiver energy management scheme where transmitter and receiver work cooperatively when either one of them needs to save power.
This scheme is best illustrated using Fig.~\ref{fig-scaled_energy}. 
Suppose the current SNR is 1 dB.
The best decoding algorithm which results in lowest energy consumption is IRRWBF algorithm and the energy consumption is about 0.1 joule.
Assume that due to battery concerns, this energy consumption is too high.
According to the graph, it is possible to achieve lower energy consumption, but it must operates at another SNR point.
Suppose the desired energy consumption is 0.05 joule, then the SNR must be 2.5 dB and the modified min-sum algorithm should be used.
SNR tells us how much signal power needs to be generated at the transmitter if the noise power is fixed.
Which means we need to increase the transmit power from 1 dB to 2.5 dB and the receiver must choose the best algorithm (MMS algorithm in this case) at that SNR regime.
Similarly, if the transmitter wants to save power, a lower SNR is assumed and the energy consumption at the receiver will increase. 
A concern about increasing transmit power is the interference to other users. 
The interference problem can be alleviated using beamforming or smart antenna~\cite{Razavilar:1999}. 
Nevertheless, when saving the receiver energy is the first priority, such as in some emergent situations, the scheme should be applied even it will interfere with other users.

\section{Conclusions}

Implementing LDPC codes on processors for software radio is a challenging problem.
In this paper, we have shown and discussed various concerns such as complexity, power consumption, and cache behaviors.
In order to do that, we proposed an efficient method to estimate power consumption.
A comparison of power consumption for different decoding algorithms were shown in this paper.

Software radio has provided us the opportunity to accomplish new power management schemes using algorithm diversity.
Two schemes were proposed: SNR-based algorithm diversity and joint transmit power and receiver energy management.
By applying the power estimation results, we were able to illustrate how these two new power control schemes work.



\bibliographystyle{IEEEtran}
\bibliography{LDPC_journal}

\begin{thebibliography}{10}
\providecommand{\url}[1]{#1}
\csname url@rmstyle\endcsname
\providecommand{\newblock}{\relax}
\providecommand{\bibinfo}[2]{#2}
\providecommand\BIBentrySTDinterwordspacing{\spaceskip=0pt\relax}
\providecommand\BIBentryALTinterwordstretchfactor{4}
\providecommand\BIBentryALTinterwordspacing{\spaceskip=\fontdimen2\font plus
\BIBentryALTinterwordstretchfactor\fontdimen3\font minus
  \fontdimen4\font\relax}
\providecommand\BIBforeignlanguage[2]{{%
\expandafter\ifx\csname l@#1\endcsname\relax
\typeout{** WARNING: IEEEtran.bst: No hyphenation pattern has been}%
\typeout{** loaded for the language `#1'. Using the pattern for}%
\typeout{** the default language instead.}%
\else
\language=\csname l@#1\endcsname
\fi
#2}}

\bibitem{Mitola:1993}
J.~Mitola, ``Software radios: Survey, critical evaluation and future
  directions,'' \emph{IEEE Aerospace and Electronic Systems Magazine}, vol.~8,
  no.~4, pp. 25--36, April 1993.

\bibitem{Lackey:1995}
{Lackey, R.I. and Upmal, D.W.}, ``Speakeasy: the military software radio,''
  \emph{IEEE Communications Magazine}, vol.~33, no.~5, pp. 56--61, May 1995.

\bibitem{Lee:2006-arch}
{Lee, C.-H. and Wolf, W.}, ``Architectures and platforms of software (defined)
  radio systems,'' \emph{International Journal of Computers and Their
  Applications}, vol.~13, no.~3, pp. 106--117, Sept. 2006.

\bibitem{Haykin:2005}
S.~Haykin, ``Cognitive radio: brain-empowered wireless communications,''
  \emph{IEEE Journal on Selected Areas in Communications}, vol.~23, no.~2, pp.
  201--220, Feb. 2005.

\bibitem{Mansour:2002}
{Mansour, M. and Shanbhag, N.}, ``Low-power {VLSI} decoder architectures for
  {LDPC} codes,'' in \emph{Proceedings of International Symposium on Low Power
  Electronics and Design}, Aug. 2002, pp. 284--289.

\bibitem{Tanner:1981}
R.~Tanner, ``A recursive approach to low complexity codes,'' \emph{IEEE
  Transactions on Information Theory}, vol.~27, no.~5, pp. 533--547, Sept.
  1981.

\bibitem{Kschischang:2001}
{Kschischang, F.R., Frey, B.J., and Loeliger, H.-A.}, ``Factor graphs and the
  sum-product algorithm,'' \emph{IEEE Transactions on Information Theory},
  vol.~47, no.~2, pp. 498--519, Feb. 2001.

\bibitem{Pearl:1988}
J.~Pearl, \emph{Probabilistic reasoning in intelligent systems}, 2nd~ed.\hskip
  1em plus 0.5em minus 0.4em\relax San Francisco, CA: Morgan Kaufmann, 1988.

\bibitem{Chiani:2000}
{Chiani, M., Conti, A., and Ventura, A.}, ``Evaluation of low-density
  parity-check codes over block fading channels,'' in \emph{Proceedings of IEEE
  International Conference on Communications}, June 2000, pp. 1183--1187.

\bibitem{ZhangT:2004}
{Zhang, T. and Parhi, K.}, ``Joint (3,k)-regular {LDPC} code and decoder/
  encoder design,'' \emph{IEEE Transactions on Signal Processing}, vol.~52,
  no.~4, pp. 1065--1079, Aug. 2004.

\bibitem{Wiber:1996}
N.~Wiberg, ``Codes and decoding on general graphs,'' {Ph.D. Dissertation},
  Linköping University, 1996.

\bibitem{Heo:2003}
J.~Heo, ``Analysis of scaling soft information on low density parity check
  codes,'' \emph{Electronics Letters}, vol.~39, no.~2, pp. 219--221, Jan. 2003.

\bibitem{Kou:2001}
{Kou, Y., Lin, S., and Fossorier, M.}, ``Low-density parity-check codes based
  on finite geometries: a rediscovery and new results,'' \emph{IEEE
  Transactions on Information Theory}, vol.~47, no.~7, pp. 2711--2736, Nov.
  2001.

\bibitem{Gallager:1962}
R.~Gallager, ``Low-density parity-check codes,'' \emph{IEEE Transactions on
  Information Theory}, vol.~8, no.~1, pp. 21--28, Jan. 1962.

\bibitem{ZhangJ:2004}
{Zhang, J. and Fossorier, M.P.C.}, ``A modified weighted bit-flipping decoding
  of low-density parity-check codes,'' \emph{IEEE Communications Letters},
  vol.~8, no.~3, pp. 165--167, March 2004.

\bibitem{Guo:2004}
{Guo, F. and Hanzo, L.}, ``Reliability ratio based weighted bit-flipping
  decoding for low-density parity-check codes,'' \emph{Electronics Letters},
  vol.~40, no.~21, pp. 1356--1358, 10 Oct. 2004.

\bibitem{Lee:2005_IRRWBF}
{Lee, C.-H. and Wolf, W.}, ``Implementation-efficient reliability ratio based
  weighted bit-flipping decoding for {LDPC} codes,'' \emph{Electronics
  Letters}, vol.~41, no.~13, pp. 755--757, 23 June 2005.

\bibitem{Sridhara:2001}
{Sridhara, D., Fuja, T., and Tanner, R.M.}, ``Low density parity check codes
  from permutation matrices,'' in \emph{Proceedings of Conference on
  Information Sciences and Systems}, March 2001.

\bibitem{Lee:2005_Power}
{Lee, C.-H. and Wolf, W.}, ``Energy/power estimation for ldpc decoders in
  software radio systems,'' in \emph{Proceedings of IEEE International Workshop
  on Signal Processing Systems Design and Implementation}, Nov. 2005, pp.
  48--53.

\bibitem{Brooks:2000}
{Brooks, D., Tiwari, V., and Martonosi, M.}, ``Wattch: a framework for
  architectural-level power analysis and optimizations,'' in \emph{Proceedings
  of International Symposium on Computer Architecture}, June 2000, pp. 83--94.

\bibitem{tensilica}
Tensilica, ``Tensilica white paper,''
  http://www.tensilica.com/pdf/Xpres\%20White\%20Paper.pdf, 2008.

\bibitem{xtensa}
Xtensa, ``Xtensa processor developers toolkit product brief,''
  http://www.tensilica.com/pdf/processor\_dev\_toolkit.pdf, 2008.

\bibitem{Atluri:2004}
{Atluri, I. and Arslan, T.}, ``Reconfigurability-power trade-offs in {Turbo}
  decoder design and implementation,'' in \emph{Proceedings of IEEE Computer
  Society Annual Symposium on VLSI}, Feb. 2004, pp. 19--21.

\bibitem{Karasawa:2000}
{Karasawa, Y., Kamiya, Y., Inoue, T., and Denno, S.}, ``Algorithm diversity in
  a software antenna,'' \emph{IEICE Transactions on Communications}, vol.
  E83-B, no.~6, pp. 1229--1236, June 2000.

\bibitem{Razavilar:1999}
{Razavilar, J., Rashid-Farrokhi, F., Liu, K.J.R.}, ``Software radio
  architecture with smart antennas: a tutorial on algorithms and complexity,''
  \emph{IEEE Journal on Selected Areas in Communications}, vol.~17, no.~4, pp.
  .662--676, April 1999.

\end{thebibliography}

%
%
%

%
%
%

\end{document}